\documentclass{ws-ijmpb}

\newcommand{\nubar}{\bar{\nu}}
\newcommand{\rhobar}{\bar{\rho}}
\newcommand{\bq}{{\bf q}}

\newcommand{\bk}{{\bf k}}
\newcommand{\br}{{\bf r}}
\newcommand{\bR}{{\bf R}}
\newcommand{\bG}{{\bf G}}
\newcommand{\Hhat}{\hat{H}}

\begin{document}

\markboth{M.\ O.\ Goerbig}
{Quantum Phases in Partially Filled LLs}
\catchline{}{}{}{}{}

\title{Quantum Phases in Partially Filled Landau Levels}
\author{\underline{M.\ O.\ Goerbig}$^{1,2,}$\footnote{E-mail: goerbig@lps.u-psud.fr}, P.\ Lederer$^2$, and C.\ Morais\ Smith$^{1,3}$}
\address{$^1$D\'epartement de Physique, Universit\'e de Fribourg, P\'erolles,
CH-1700 Fribourg, Switzerland.\\
$^2$Laboratoire de Physique des Solides, B\^at. 510 (associ\'e au CNRS) 
91405 Orsay cedex, France.\\
$^3$Institute for Theoretical Physics, Utrecht University, Leuvenlaan 4,\\
3584 CE Utrecht, The Netherlands.}

\maketitle

\begin{history}
\received{25 June 2004}
\revised{20 August 2004}
\end{history}

\begin{abstract}
We compare the energies of different electron solids, such as bubble crystals
with triangular and square symmetry
and stripe phases, to those of correlated quantum liquids 
in partially filled intermediate Landau levels. Multiple
transitions between these phases when varying the filling of the top-most 
partially filled Landau level explain the observed reentrance of the
integer quantum Hall effect. The phase transitions are identified as
first-order. This leads to a variety of measurable phenomena
such as the phase coexistence between a Wigner crystal and a two-electron 
bubble phase in a Landau-level filling-factor range 
$4.15\lesssim \nu \lesssim 4.26$, which has recently been observed in 
transport measurements under micro-wave irradiation.
\end{abstract}


\section{Introduction}

The most prominent phenomena in two-dimensional electron systems 
in a perpendicular magnetic field $B$ are
the integer and fractional quantum Hall effects (IQHE and
FQHE, respectively). In spite of the similarity
between the two effects, their origin is different: on the one hand,
the IQHE is a manifestation of
the energy quantization of electrons (mass $m$ and
charge $-e$) in highly degenerate Landau levels (LLs), with a level 
separation of $\hbar eB/m$.
The ratio $\nu=n_{el}/n_B$ between the electronic density
$n_{el}$ and the density of states per level, $n_B=B/(h/e)$, determines the
filling of the LLs, and the IQHE occurs if $\nu=N$, with integral
$N$. The signature of this effect is a plateau in the Hall resistance,
accompanied by a vanishing longitudinal resistance.
On the other hand, the FQHE is due to strongly correlated
quantum liquids  formed by the electrons in a partially
filled LL and 
occurs at some of the ``magical'' filling factors $\nu=p/(2ps+1)$
[and at their particle-hole symmetric fillings $\nu=1-p/(2ps+1)$],
with integral $s$ and $p$. Also in the first excited LL, fractional
quantum Hall states have been observed at $\nubar=1/3,2/3,1/5$, and $4/5$,
where $\nubar=\nu-N$ denotes the filling of the topmost level. 

In higher LLs, the strong Coulomb repulsion between electrons in the
partially filled level may lead to phases different from quantum
liquids: calculations in the Hartree-Fock approximation have revealed the
existence of electron-solid phases, such as stripes around
$\nubar=1/2$ and bubble crystals with varying electron number per
lattice site at $\nubar<1/2$.\cite{FKS,moessner} A stripe phase has indeed been
observed in transport measurements, which show a large anisotropy in
the longitudinal magneto-resistance around
$\nu=9/2,11/2,13/2,\ldots$\cite{exp1} Eisenstein {\sl et al.} have furthermore
measured a non-monotonic behavior of the Hall resistance in the first
excited LL $n=1$:\cite{exp3} the FQHE at $\nubar=1/3$ and $1/5$ is
surrounded by pinned electron-solid phases, which are insulating and thus
cause an integer
quantization of the Hall resistance, as for the neighboring IQHE. This
reentrant IQHE is reminiscent of an effect observed before in the second
excited LL.\cite{exp2} We
have shown that the effect may be understood in terms of an alternation
between quantum-liquid and electron-solid phases when varying the
filling of the topmost LL.\cite{goerbig05} Here, we furthermore investigate 
bubble crystals with different symmetry. Whereas the quantum-liquid phases
are favored at $\nubar=1/(2s+1)$, at $\nubar\neq1/(2s+1)$
quasi-particles are excited and raise the energy of the quantum
liquids above that of the competing electron solids. 

\section{Energy Calculation for the Different Phases}

In order to describe the low-energy degrees of freedom, which, at non-zero
values of the partial filling factor $\nubar$, consist of intra-LL excitations,
we adopt a model of spin-polarized electrons,
\begin{equation}
\label{equ001}
\Hhat=\frac{1}{2}\sum_{\bq}v_n(q)\rhobar(-\bq)\rhobar(\bq),\qquad
{\rm with}~~~~v_n(q)=\frac{2\pi e^2}{\epsilon q}\left[F_n(q)\right]^2,
\end{equation}
where only the components of the density operator in the $n$-th LL are taken
into account, $\rho_n(\bq)=F_n(q)\rhobar(\bq)$.\footnote{We use a
  system of units, in which the magnetic length $l_B=\sqrt{\hbar/eB}\equiv1$}
The LL form factor $F_n(q)=L_n(q^2/2)\exp(-q^2/4)$ is given in terms of
Laguerre polynomials $L_n(x)$, and $\epsilon$ is the dielectric constant.
The quantum-mechanical properties of the model are revealed by the unusual
commutation relations for the projected density operators,
$[\rhobar(\bq),\rhobar(\bk)]=2i\sin\left[(\bq\times\bk)_z/2\right]
\rhobar(\bq+\bk)$.
This model allows for a common description of all LLs.

The electron-solid phases are characterized by an order parameter
$\Delta(\bq)$, which 
determines the density profile of the phase, given by the local filling
factor $\nubar(\br)$ and the area $A$, 
$\Delta(\bq)\equiv\langle\rhobar(\bq)\rangle/n_BA=
\int d^2r\nubar(\br)\exp(i\bq\cdot\br)/A$. The cohesive energy of the
electron-solid phases becomes in the Hartree-Fock 
approximation\cite{FKS,moessner,goerbig05}
\begin{equation}
\label{equ004}
E_{coh}^{sol}(n;\nubar)=\frac{n_B}{2\nubar}\sum_{\bq}u_n^{HF}(q)
|\Delta(\bq)|^2,
\end{equation}
where the Hartree-Fock potential $u_n^{HF}({\bf q})$ takes into account 
quantum-mechanical exchange effects. 

The bubble crystal with an arbitrary lattice symmetry is characterized by the
local filling factor $\nubar(\br)=\Theta(r_B-|\br-\bR_j|)$, where $\Theta(x)$ 
is the step function, and $\bR_j$ are the lattice vectors. The area of the
primitive cell $A_{pc}=2\pi M/\nubar$ is determined by the partial 
filling factor and the bubble radius $r_B=\sqrt{2M}$ containing $M$ 
electrons. The order parameter of the bubble crystal 
$$\Delta_M^B(\bq)=\frac{2\pi\sqrt{2M}}{Aq}J_1(q\sqrt{2M})\sum_j 
e^{i\bq\cdot{\bf R}_j}$$
yields the cohesive energy
\begin{equation}
\label{equ005}
E_{coh}^{B}(n;M,\nubar)=\frac{n_B\nubar}{M}\sum_{\bG_l\neq 0} u_n^{HF}(\bG_l)
\frac{J_1^2(\sqrt{2M}|{\bf G}_l|)}{|{\bf G}_l|^2},
\end{equation}
where the lattice symmetry is specified only by the reciprocal lattice 
vectors ${\bf G}_l$. 

In the case of stripes with width $a$ oriented parallel to the 
$y$-direction, the ansatz $\nubar(\br)=\Theta(a/2-|x-x_j|)$ leads to the
order parameter,
$$\Delta^S(\bq)=\frac{2}{L_x}\delta_{q_y,0}
\frac{\sin\left(q_x\Lambda_S\nubar/2\right)}{q_x}\sum_je^{iq_xj\Lambda_S},$$
where $\Lambda_S=a/\nubar$ is the stripe periodicity. This yields the 
cohesive energy
\begin{equation}
\label{equ006}
E_{coh}^{S}(n;\Lambda_S,\nubar)=\frac{n_B}{2\pi^2\nubar}\sum_{l\neq0}
u_n^{HF}\left(q=\frac{2\pi}{\Lambda_S}l\right)\frac{\sin^2(\pi\nubar l)}{l^2},
\end{equation}
which is to be minimized with respect to the variational parameter $\Lambda_S$.

The quantum-liquid phases, which we investigate here,
may not be characterized by an order parameter, but they are described 
by Laughlin's wavefunctions.\cite{laughlin}
Their cohesive energy is given in
terms of Haldane's pseudopotentials,\cite{haldane} 
$V_{2m+1}^{n}=(2\pi/A)\sum_{\bq}v_n(q)L_{2m+1}(q^2)\exp(-q^2/2)$,
\begin{equation}
\label{equ007}
E_{coh}^{q-l}(n;s,\nubar)=\frac{\nubar}{\pi}\sum_{m=0}^{\infty}c_{2m+1}^s
V_{2m+1}^n+[\nubar(2s+1)-1]\Delta^n(s),
\end{equation}
where the expansion coefficients $c_{2m+1}^s$ specify the Laughlin 
wavefunction. The second term in Eq. (\ref{equ007}) takes 
into account the energies $\Delta^n(s)$ of the excited quasi-particles 
of charge $1/(2s+1)$ [at $\nubar>1/(2s+1)$] and quasi-holes of charge 
$-1/(2s+1)$ [at $\nubar<1/(2s+1)$], in units of the electronic charge.
They may be calculated analytically in the 
Hamiltonian theory of the FQHE, established by Murthy and Shankar.\cite{MS}

\section{Results}

Here, we concentrate on some aspects of the phases in the first and second
excited LLs, $n=1$ and $n=2$, respectively. A more detailed discussion, 
including a quantitative study of the role of impurities, may be
found in Ref.~\refcite{goerbig05}.

Fig. 1(a) shows the energies for different electronic phases in
$n=1$. The quantum-liquid phases are favored around $\nubar=1/3$ and $1/5$,
whereas in an intermediate range, $0.23<\nubar<0.3$, a Wigner crystal (WC,
$M=1$) has a lower energy. Because the Wigner is pinned by impurities, one 
observes an integer quantization of the Hall resistance in this range, 
whereas one finds the FQHE around
$\nubar=1/3$ and $1/5$.\cite{exp3} Above $\nubar\sim0.38$, the FQHE disappears
because the quantum liquid has a higher energy than a two-electron bubble
crystal, which competes with a stripe phase. The latter has a lower energy as
one approaches half-filling. Experimentally, however, an anisotropic 
longitudinal resistance, which is the signature of stripe phases,\cite{exp1}
has only been observed in a tilted magnetic field.\cite{exp4} Notice that 
non-Laughlin-type quantum liquids, which are not considered in our energy 
investigations, also compete in this filling-factor range. At $\nu=5/2$, 
{\it e.g.}, a Pfaffian state, which is a special case ($k=2$) of the 
parafermionic ones at $\nubar=k/(2+k)$ with integral $k$, gives rise to a 
FQHE.\cite{pfaff} A recently observed FQHE\cite{xia} at $\nubar=2/5$ in $n=1$
is likely to be a parafermionic hole state with $k=3$. Also the nature of the
FQHE state at $\nubar=1/3$ remains controversial because numerical studies
on a few number of electrons indicate a rather small overlap with a 
Laughlin-type state.\cite{nu13} Although we consider only such Laughlin-type 
states here, it cannot be ruled out that other
quantum-liquid phases have a lower energy and are responsible for the FQHE at 
these fillings.

Our energy calculations suggest that 
quantum-liquid phases may also be found below $\nubar=1/5$ in the absence of
impurities. However, the energy of the WC is lowered by impurities, due to the
deformation of its lattice structure. This effect is most relevant at small 
$\nubar$, and the FQHE is therefore unstable in this 
limit,\cite{goerbig05} where one observes the IQHE.\cite{exp3} The energies
for the bubble crystals are shown both for the case of a triangular (continuous
lines) and a square lattice symmetry (broken lines). The energy difference 
between these two cases is extremely tiny (on the order of $1\%$). From
classical considerations, one would expect that a triangular lattice has a
lower energy than a square lattice.\cite{bonsall} Our energy calculations
indicate that this is correct in the low-$\nubar$ limit, whereas at larger 
densities a WC with square-lattice symmetry has a lower energy than 
the triangular one. A similar behavior is found for the two-electron bubble
crystal. However, this change of symmetry occurs at filling-factor values, 
where other phases have a lower energy; the square-lattice symmetry of the WC, 
{\it e.g}, is favored only above $\nubar\sim0.3$, where quantum-liquid, 
two-electron bubble, and stripe phases are the ground state.

\begin{figure}[htbp]
\centerline{\psfig{file=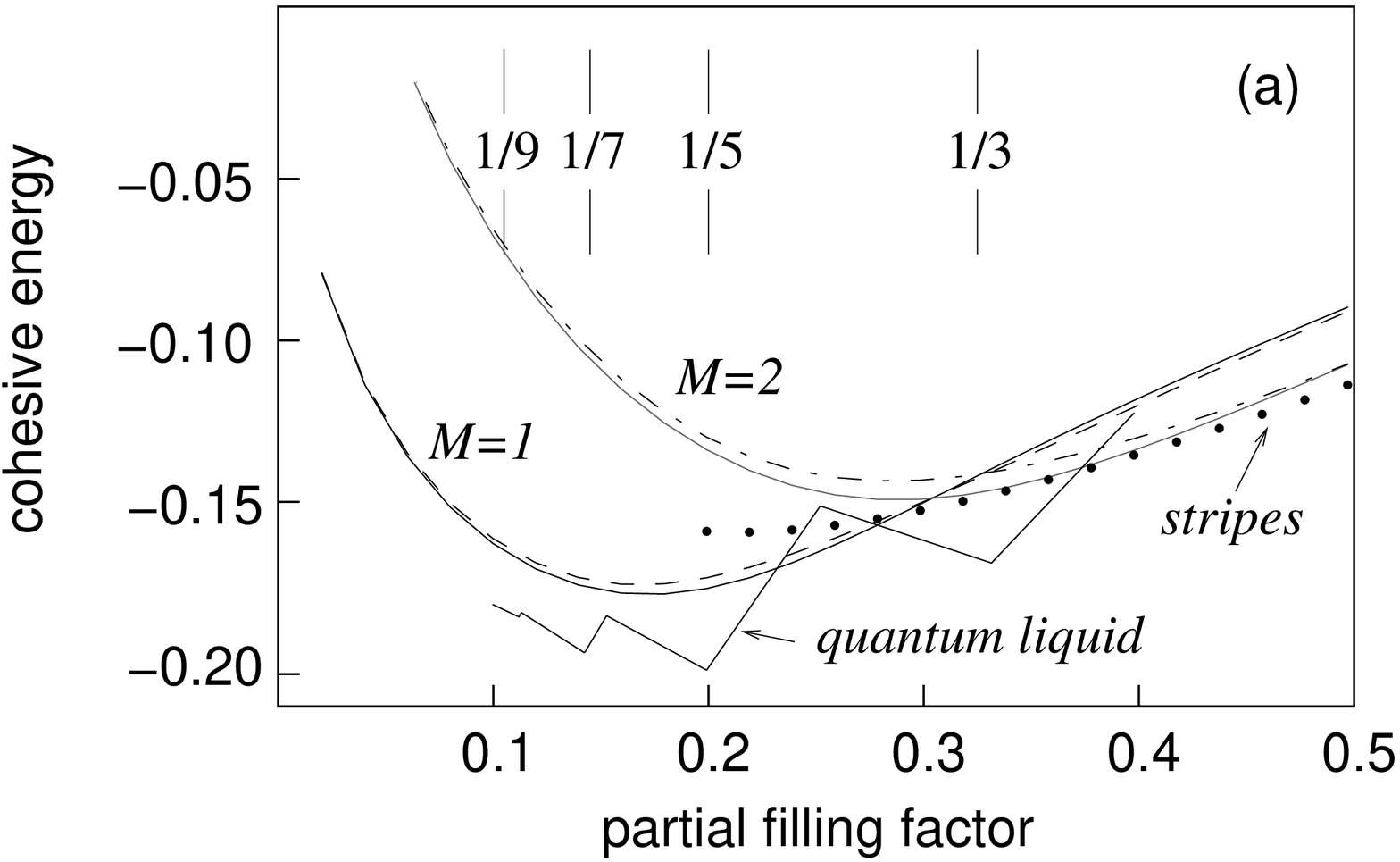,width=6.5cm}
\psfig{file=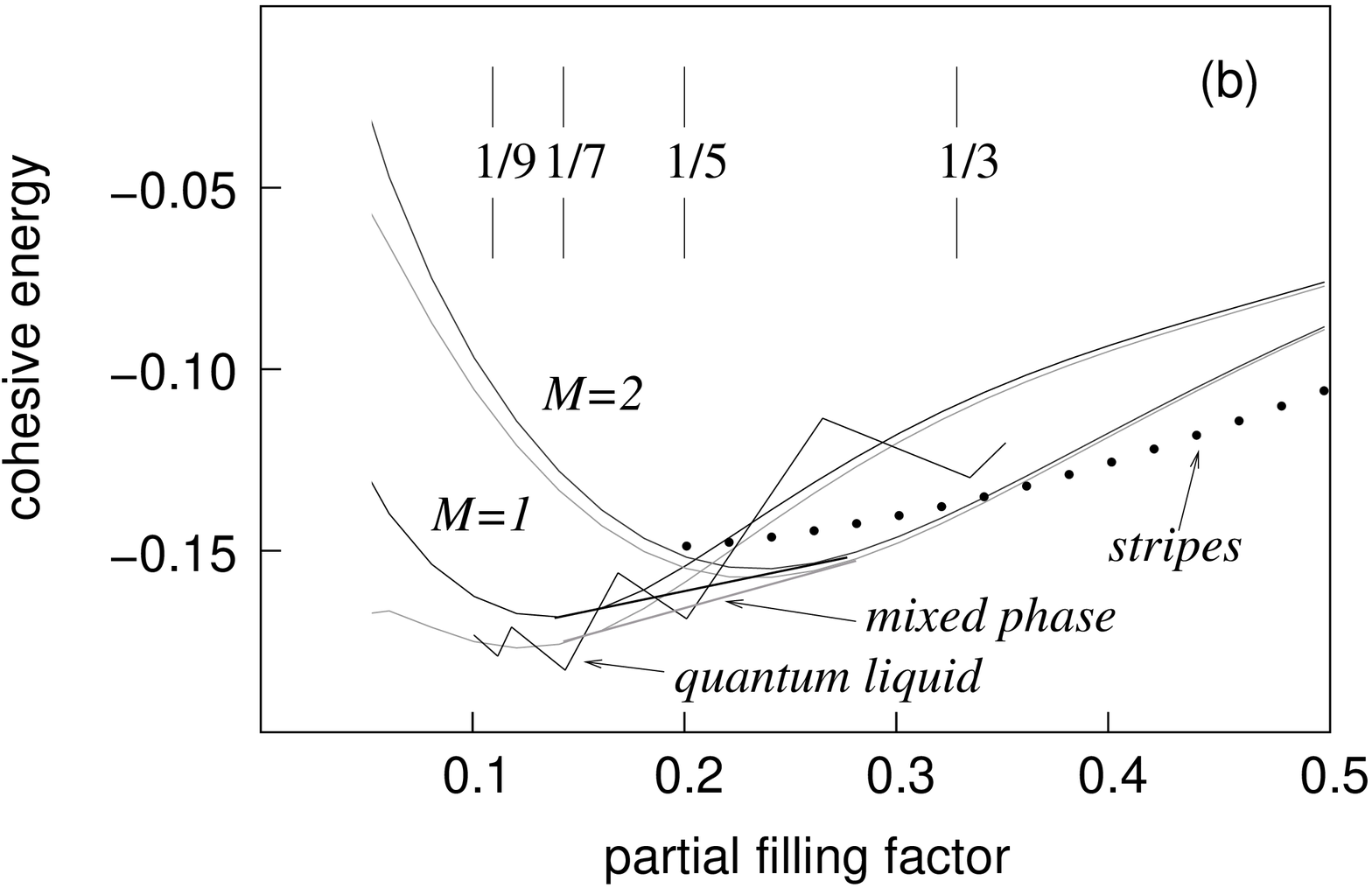,width=6.5cm}}
\vspace*{8pt}
\label{fig01}
\caption{Cohesive energies of the different phases, in units of
$e^2/\epsilon l_B$. (a): phases in $n=1$. For the bubble phases, both 
the triangular (continuous lines) and the square crystal (broken lines) are 
shown. (b): phases in $n=2$. The gray lines indicate the 
bubble-crystal energies in the presence of an impurity potential, and the 
tangents represent a mixed phase.}
\end{figure}

The energy results for $n=2$ are shown in Fig. 1(b). In contrast to
$n=1$, a quantum liquid is unstable around $\nubar=1/3$, where a two-electron
bubble crystal has the lowest energy. Our energy calculations suggest that
a FQHE might be found around $\nubar=1/5$ or $1/7$. Note, however, that the 
energies of the quantum-liquid phases are very close to that of the WC and,
in the case of $\nubar=1/5$, to a mixed phase of a WC and a two-electron bubble
crystal, which is represented by the tangent. It is therefore not
clear whether the quantum liquid remains stable in the presence of impurities,
which lower the energy of the crystal phases, as shown by the gray curves.
They have been calculated for an impurity strength $V_0/\xi=0.005 e^2/\epsilon
l_B^2$, where $V_0$ is the characteristic energy of a short-range Gaussian 
potential with correlation length $\xi$.\cite{goerbig05} 
Experimentally, a small maximum in the longitudinal resistance around 
$\nubar=1/5$ indicates an incipient melting of a crystal phase.\cite{exp2} 
This feature has recently been studied in more detail by Gervais 
{\it et al.},\cite{gervais} who found that the maximum, 
which decreases when lowering the temperature $T$, splits into two peaks 
separated by a small local minimum precisely
at $\nubar=1/5$ with increasing $T$. A reminiscent 
$T$-dependent effect has been observed in the WC regime in the lowest 
LL.\cite{pan1} Even if this effect may indicate a quantum-liquid ground state
in extremely pure samples, it may also be understood in different terms:
whereas the crystal, which in this scenario remains 
the $T=0$ ground state, melts at rather low $T$ (on the order
of the energy difference between the WC and the quantum-liquid phase), the 
quantum coherence of the liquid displaying FQHE features is only destroyed at
higher $T$.\cite{goerbig02} 

\section{Phase Transitions}

Our energy calculations suggest that the transitions between the 
different phases are {\it first-order}. The first-order phase 
transitions between the quantum-liquid and the insulating bubble crystals may
cause a hysteretical behavior in the Hall resistance around the transition
points, which, to the knowledge of the authors, has not been
reported yet. Also the phase transitions between 
bubble crystals with different $M$ per site are first-order, in agreement
with time-dependent Hartree-Fock calculations by C\^ot\'e 
{\it et al.}\cite{cote} This leads to a phase coexistence, or a mixed phase,
around the transition points in a filling-factor range, which is described
by a tangent on the energy curves, {\it e.g.} at $0.15\lesssim \nubar \lesssim 
0.26$ in $n=2$ [c.f. Fig 1(b)]. Experimentally, there is evidence for
such a mixed phase, which is revealed by a double-peak structure in transport
measurements under micro-wave irradiation, recently performed by Lewis 
{\it et al.}\cite{lewis} 

\section{Conclusions}

In conclusion, we have performed energy calculations for competing quantum 
phases in intermediate LLs. An alternation between insulating electron-solid
and quantum-liquid phases, which display the FQHE, is at the origin of the 
observed reentrance of the IQHE in $n=1$ and $n=2$.\cite{exp2,exp3} The 
transitions between the different phases are found to be first-order and may
lead to a variety of observable phenomena. In the case of transitions between 
bubble crystals with different electron number per site, a phase coexistence
is expected.\cite{goerbig05} This scenario is supported by recent micro-wave 
experiments, in which a double-peak structure has been observed in the 
longitudinal conductivity in a filling-factor range $0.16\lesssim \nubar
\lesssim 0.28$,\cite{lewis} in good agreement our theoretical 
investigations.\cite{goerbig05}

\section*{Acknowledgments}

We acknowledge fruitful discussions with K.\ Borejsza, J.\ P.\ Eisenstein, 
T.\ Giamarchi, R.\ Lewis, R.\ Moessner, and 
S.\ Scheidl. This work was supported by the Swiss National Foundation for 
Scientific Research under grant No.~620-62868.00.


\begin{thebibliography}{0}

\bibitem{FKS}A. A. Koulakov, M. M. Fogler, and B. I. Shklovskii, 
{\it Phys. Rev. Lett.} {\bf 76}, 499 (1996);
M. M. Fogler, A. A. Koulakov, and B. I. Shklovskii, {\it Phys. Rev.}
{\bf B54}, 1853 (1996); M. M. Fogler and  A. A. Koulakov,
{\it ibid.} {\bf B55}, 9326 (1997).

\bibitem{moessner}R. Moessner and J. T. Chalker, 
{\it Phys. Rev.} {\bf B54}, 5006 (1996).

\bibitem{exp1} M. P. Lilly {\it et al.}, {\it Phys. Rev. Lett.} {\bf 82},
  394 (1999);  R. R. Du {\it et al.}, {\it Solid State Commun.} {\bf 109}, 
  389 (1999).

\bibitem{exp3}J. P. Eisenstein {\it et al.}, {\it Phys. Rev. Lett.} {\bf 88}, 
076801 (2002). 

\bibitem{exp2}K. B. Cooper {\it et al.}, {\it Phys. Rev.} {\bf B60},
11285 (1999).

\bibitem{goerbig05}M. O. Goerbig, P. Lederer, and C. Morais Smith, 
{\it Phys. Rev.} {\bf B68}, 241302 (2003); {\it Phys. Rev.} {\bf B69}, 
115327 (2004).

\bibitem{laughlin}R. B. Laughlin, {\it Phys. Rev. Lett.} {\bf 50}, 1395
  (1983).

\bibitem{haldane}F. D. Haldane, {\it Phys. Rev. Lett.} {\bf 51}, 605 (1983).

\bibitem{MS}G. Murthy and R. Shankar, {\it Rev. Mod. Phys.} {\bf 75}, 
1101 (2003).

\bibitem{exp4}W. Pan {\it et al.}, {\it Phys. Rev. Lett.} {\bf 83}, 820 (1999);
M. P. Lilly {\it et al.}, {\sl ibid.} {\bf 83}, 824 (1999).

\bibitem{pfaff}G. Moore and N. Read, {\it Nucl. Phys.} {\bf B360}, 362 
(1991); N. Read and E. Rezayi, {\it Phys. Rev.} {\bf B59}, 8084 (1999).

\bibitem{xia}J. S. Xia {\it et al.}, preprint: cond-mat/0406724.

\bibitem{nu13}N. d'Ambrumenil and A. M. Reynolds, {\it J. Phys.} {\bf C21},
119 (1988); A. Wojs, {\it Phys. Rev.} {\bf B63}, 125312 (2001).

\bibitem{bonsall}L. Bonsall and A. A. Maradudin, {\it Phys. Rev.} {\bf B15}, 
1959 (1977).

\bibitem{gervais}G. Gervais {\it et al.}, preprint: cond-mat/0402169.

\bibitem{pan1}W.\ Pan {\it et al.}, {\it Phys. Rev. Lett.} {\bf 88}, 
176802 (2002).

\bibitem{goerbig02}M. O. Goerbig and C. Morais Smith, {\it Europhys. Lett.} 
{\bf 63}, 736 (2003).

\bibitem{cote}R. C\^ot\'e {\it et al.}, {\it Phys. Rev.} {\bf B68}, 155327 
(2003).

\bibitem{lewis}R. M. Lewis {\it et al.}, preprint: cond-mat/0401462.

\end{thebibliography}
\end{document}